\documentclass[aps, prl, reprint, superscriptaddress]{revtex4-2}
\usepackage{amsmath,amssymb}
\usepackage{graphicx}
\usepackage{dcolumn}
\usepackage{bm}
\usepackage{braket}
\usepackage{verbatim}
\usepackage{bbold}
\usepackage{color}
\usepackage{xcolor}
\usepackage{relsize}
\usepackage{slashed}
\usepackage{amsthm}
\usepackage[colorlinks=true, urlcolor=blue, urlbordercolor={0 1 1}]{hyperref}

\begin{document}

\title{Self-doped Crystal from Preempted Band-inversion Transitions}
\author{Jiechao~Feng}
\affiliation{Graduate Group in Applied Science \& Technology, University of California, Berkeley, CA 94720, USA}
\author{Zhaoyu~Han}
\affiliation{Department of Physics, Harvard University, Cambridge, Massachusetts 02138, USA}
\author{Michael~P.~Zaletel}
\affiliation{Department of Physics, University of California, Berkeley, CA 94720, USA}
\author{Zhihuan~Dong}
\affiliation{Department of Physics, University of California, Berkeley, CA 94720, USA}

\date{\today}

\begin{abstract}
Recent experiments in rhombohedral graphene  find evidence for a ``self-doped’’ Wigner crystal  (SDC) in which a slightly incommensurate Wigner crystal (WC) coexists with a small Fermi sea. ~\cite{han2026evidence}.
We provide non-perturbative arguments that such SDCs  generically arise from preempted band-inversion transitions between commensurate crystals, which  motivates simple band-theory criteria for their appearance. 
Self-consistent Hartree-Fock calculations establish the existence of a SDC consistent with this mechanism in both the $\lambda$-jellium model and rhombohedral pentalayer graphene (R5G). 
In the $\lambda$-jellium model, we identify a SDC phase located between a ``halo''-WC and an anomalous Hall crystal (AHC), which would otherwise be connected via a Dirac transition when pinned to commensuration; this contrasts with the WC-AHC transition, which we show cannot be connected by a continuous transition due to a mismatch of symmetry indices. In R5G, we predict a SDC phase located between a WC and a ``disqualified'' halo anomalous Hall crystal.  We discuss in general how the Berry curvature distribution in the parent band affects the appearance of SDC, revealing a novel role of quantum geometry in inducing exotic quantum phases of matter.
\end{abstract}

\maketitle

\paragraph*{Introduction---}
The Wigner crystal~(WC) is one of the 
paradigmatic examples of interaction-driven order in a two-dimensional electron system~\cite{PhysRev.46.1002}. In the low-density limit, Coulomb repulsion dominates over kinetic energy and drives electrons into a crystalline arrangement, producing an insulating state with one electron per unit cell. While the WC is stabilized in the dilute or strong coupling limit, the broader problem of electron crystallization in the intermediate coupling regime has long remained a challenge. In particular, already in the conventional jellium setting, it was proposed that the electron filling in the crystalline state need not remain exactly commensurate: an electron crystal may self-dope and develop a small density of itinerant carriers while its crystalline long-range order persists, breaking continuous translational symmetry. Such incommensurate or self-doped crystals (SDC) have been discussed theoretically~\cite{PhysRevB.70.155114,PhysRevB.77.085104,PhysRevB.109.235130,PhysRevB.100.155132} for decades and examined numerically~\cite{katomeris2003andreev,nemeth2003andreev,PhysRevLett.86.492,PhysRevLett.94.046801}, but the conditions under which they are stabilized have remained inconclusive, and an experimental verification has remained elusive. 

Recent years have seen increasingly direct evidence for electron crystallization in clean two-dimensional systems~\cite{lu2024fractional, xiang2025imaging, ge2025visualizing,tsui2024direct,li2021imaging}.
A particularly striking case is provided by rhombohedral multilayer graphene (RMG). This class of system is known to have strong and tunable quantum geometry, which extends the experimental study of 2D electron gas beyond the conventional jellium setting. In earlier studies, a zoo of exotic quantum phases, including chiral superconductivity~\cite{han2025signatures}, quantum anomalous Hall~(QAH), and fractional quantum anomalous Hall~(FQAH) states~\cite{han2024correlated,liu2024spontaneous,zhou2024layer,sha2024observation,han2024large,lu2025extended,choi2025superconductivity,xie2025tunable,lu2014quantum,ding2026spin} have been demonstrated in the same system hosting a WC phase, highlighting the crucial role and rich and striking consequences of quantum geometry.
Remarkably, a recent experiment in RMG~\cite{han2026evidence} reveals a phase with 
features of SDC, namely a metallic phase developing next to a WC phase, showing a small and negative Hall density being approximately $1/10$ of the total electron density. Understanding this phase is important not only for its own sake. Earlier experiments also revealed a nearby superconductivity phase (SC) whose microscopic origin remains unresolved~\cite{han2025signatures}. The potential connection between SDC and SC implies that clarifying the mechanism that stabilizes the former may be essential for understanding the latter.

In this work, we point out that SDC can be stabilized as a generic consequence of criticality between undoped (commensurate) crystals. Namely, when two undoped crystals are connected through a band inversion with a critical Dirac theory, such a direct continuous transition is generically preempted by an intermediate SDC if not becoming first-order. The picture is simple and non-perturbative: at the would-be band-inversion symmetry, the chemical potential generically doesn't cross the Dirac point, so that Fermi pocket(s) are expected to be induced by adjustment of the lattice constant.
This mechanism also reveals a close connection to quantum geometry, as the relevant band inversion is tied to a criticality between topologically distinct crystalline states. Specifically, we point out that the symmetry indices characterizing the crystalline-symmetry-protected topological order~\cite{fang2012bulk, po2017symmetry, song2017topological, thorngren2018gauging, song2020real}
of undoped crystals provide crucial constraints on the existence of continuous band-inversion transitions that can induce SDCs. Remarkably, the two major principles in our framework: (1) Connection between SDC and band-inversion transition, and (2) Connection between band-inversion and SPT classification of crystals, are both beyond mean-field theory~\cite{song2017topological, thorngren2018gauging}.

We establish the existence of SDC phase from this mechanism in two examples: the $\lambda$-jellium model~\cite{x53d-12s6} and a realistic model for rhombohedral pentalayer graphene (R5G)~\cite{PhysRevB.82.035409,PhysRevB.88.075408,PhysRevLett.133.206503}.
In the $\lambda$-jellium model, where the quantum geometry and coupling strength are independently tunable, we find a SDC at the transition between AHC and halo Wigner crystal (hWC), which is the only transition allowed to be continuous by $C_6$ symmetry. 
In R5G, we theoretically predict the structure of the phase diagram from non-interacting band structure and propose a halo anomalous Hall crystal (hAHC) as a parent undoped crystal state for the SDC. This parent hAHC is distinct from the anomalous Hall crystal (AHC) discussed previously~\cite{PhysRevLett.133.206503,PhysRevLett.133.206504,PhysRevLett.133.206502} by their $C_3$ symmetry indicators. Then we show on symmetry grounds that only the WC-hAHC transition is allowed to be continuous, which explains why the self-doped crystal appears only in the experimentally relevant part of the phase diagram. Our theoretical insights are confirmed by numerics using self-consistent Hartree-Fock~(SCHF) calculations (see Appendix~\ref{app:SCHF} for details.) 
Taken together, these results provide a unified framework for understanding when and why SDCs arise. 

\paragraph*{Thermodynamic criterion for SDC---}
\label{sec:intermediate}
We are interested in the thermodynamic criterion for a commensurate crystal to give way to SDC. We provide a pedagogical explanation here, while a more rigorous derivation is detailed in the End Matter. To get some intuition, imagine deforming an electron system at fixed electron density $n_e$ from a commensurate crystal to SDC. This is achieved through two steps: (1)~slightly change the crystal density $n_c$ (number of unit cells per unit area), while keeping the one-electron-per-site constraint of the crystal, which changes the energy by a chemical potential $\mu_C = \left(\frac{\partial \epsilon}{\partial n_c}\right)_{\nu=1}$, where $\nu\equiv \frac{n_e}{n_c}$ is the electron filling of the crystal and $\epsilon$ is energy per unit area. This step also changes the electron density $n_e$, so it requires that we further (2)~electron/hole dope the compressed crystal to compensate for the change in electron density, which reduces the energy by the conduction band bottom $\xi_+$/ the valence band top $\xi_-$. The SDC is favored when the combined action of the two steps reduces the energy, that is
\begin{align}
    \mu_C<\xi_- &\Rightarrow \text{hole-doped SDC} \nonumber\\
    \mu_C>\xi_+ &\Rightarrow \text{electron-doped SDC}
    \label{criterion}
\end{align}
where we have generalized to the $e$-doped case, and $\xi_+$ is the energy of the doped electron, which, at the mean-field level, corresponds to the conduction band edge. Deep in a WC where the band gap is large, $\mu_C$ is sandwiched between the two band edges, so that the crystalline order is pinned to be commensurate with density by the quasi-particle charge gap. 

\begin{figure}[t]
\centering
\includegraphics[width=\columnwidth]{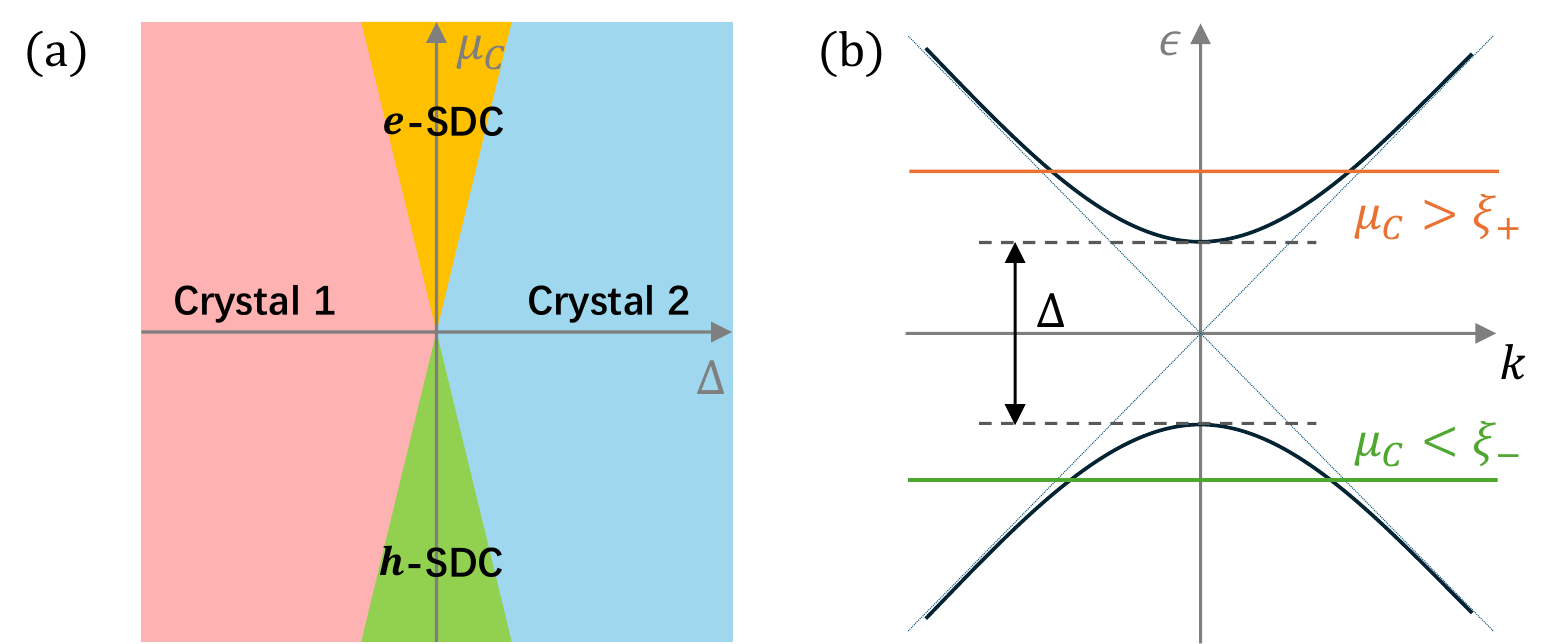}
\caption{(a)~Schematic phase diagram for SDC preempting a band-inversion transition between two undoped (commensurate) crystals, where $\Delta=0$. (b)~Self-doping happens when chemical potential moves beyond the band edges, which is generically expected as the band gap closes, approaching the critical point.}
\label{mu_Delta_pd}
\end{figure}

One immediate implication from reformulating the SDC criterion as Eq.~\eqref{criterion} is that a SDC is 
expected to preempt a band-inversion transition between two undoped crystals: as the gap closes, $\mu_C$ will not coincide with the Dirac point energy without fine-tuning, and generically lies within a band. In Fig.~\ref{mu_Delta_pd} we show a schematic phase diagram. The above analyses are non-perturbative and general.

A natural question is when such a band-inversion transition is allowed. We point out the importance of the symmetry indicators, and further show that symmetry indicators of undoped crystal phases are predictable from non-interacting band structure by studying two examples: $\lambda$-jellium model \cite{x53d-12s6}, and a realistic model of R5G~\cite{PhysRevB.82.035409,PhysRevB.88.075408,PhysRevLett.133.206503}, which is strongly relevant to a recent experiment~\cite{han2026evidence}. Surprisingly, despite having very different band structures and quantum geometries, the two problems appear to be quite similar in both the competition between crystalline order and the stability of SDC. In fact, most of our understanding of $\lambda$-jellium can be applied to the realistic R5G model with minor changes.

\begin{figure}[h]
\centering
\includegraphics[width=\columnwidth]{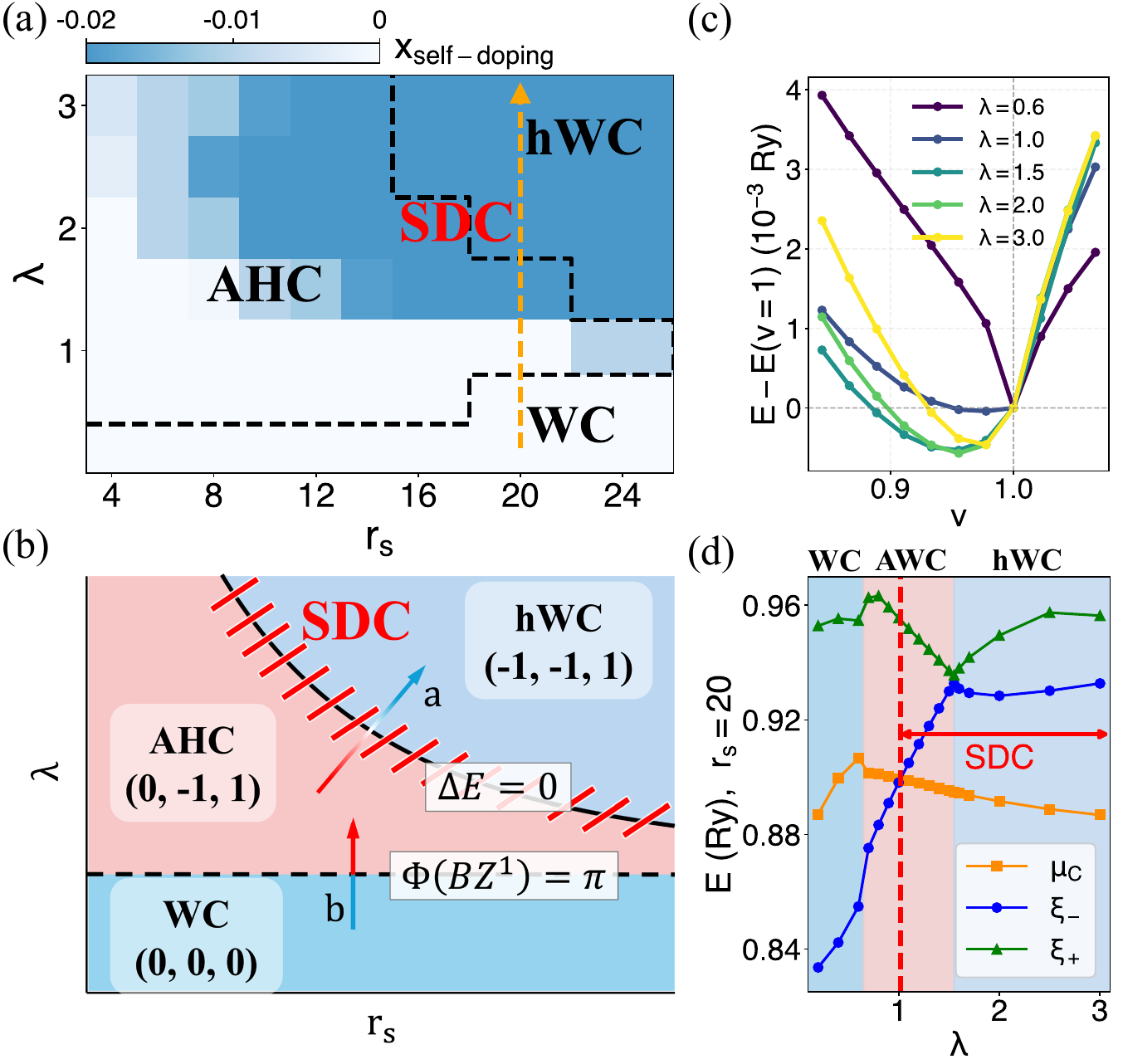}
\caption{
Phase diagram for $\lambda$-jellium model.
(a) SCHF phase diagram enforcing commensurate condition $\nu=1$. The black dashed line delineates the phases at commensurate filling $\nu=1$. The blue colored region is where the criterion Eq.~\eqref{criterion} is fulfilled, and spontaneous hole doping occurs. The color density represents the fraction of self-doping $x_{\rm self-doping}$, defined in Eq.~\eqref{x_def}. (b) Schematic phase diagram in $\lambda-r_s$ space, concerning only competition between commensurate crystalline orders. Numbers in brackets correspond to $C_6$ symmetry indicators at three high symmetry points ($l_\Gamma, l_K, l_M$). The red hatched region is where SDC is expected from our theory.
(c) Ground state energy difference relative to the commensurate ($\nu =1$) state within SCHF as a function of $\nu$ at $r_s=20$.  (d) The evolution of chemical potential $\mu_C$, valence band top $\xi_-$, and conduction band bottom $\xi_+$ along the gold arrow in panel (a) at $r_s=20$. Background colors distinguish the three commensurate crystalline phases before considering self-doping. Hole self-doping sets in above $\lambda\sim1.0$ (red dashed line).}
\label{LJ_schematic}
\end{figure}

\paragraph*{Warmup: SCHF results on SDC in $\lambda$-jellium model ---}
First, using a toy model, we demonstrate that our analytical framework fully explains the location of the SDC in the SCHF phase diagram.
The $\lambda$-jellium model was proposed as a toy two-band model to realize independently tunable Berry curvature and dispersion~\cite{x53d-12s6} and admits systematic generalizations to multi-band cases~\cite{han2025,tan2025, Desrochers2026}. Its phase diagram, restricted to commensurate crystalline orders, has been studied using SCHF~\cite{x53d-12s6, Desrochers2026} and QMC~\cite{valenti2025quantum}. In our SCHF study, we will first work with undoped (commensurate) crystal ansatzes (i.e. fixing $\nu=1$), and then use the criterion Eq.~\eqref{criterion} to address the instability toward SDC, which is beyond the scope of previous works.  Comparing $\mu_{C}$ and HF band edges, we determine the instability towards SDC, and also estimate the self-doping fraction, i.e. the fraction of electrons forming the itinerant fermi surface 
\begin{equation}
    x_{\rm self-doping}=\frac{n_{< \mu_C}}{n_e}-1 \label{x_def}
\end{equation}
where $n_{<\mu_C}$ is the density of electron states below $\mu_C$ in the Hartree-Fock (HF) bands. The result is shown in Fig.~\ref{LJ_schematic}(a), overlayed with the phase boundaries between commensurate crystals. The effectiveness of SDC criterion Eq.~\eqref{criterion} is further supported by a direct calculation that relaxes the commensurate ($\nu=1$) constraint, shown in Fig.~\ref{LJ_schematic}(c).

The commensurate crystal phase diagram is divided into three phases: WC at small $\lambda$, AHC at higher $\lambda$ small $r_s$ and hWC at higher $\lambda$ and large $r_s$.  The SDC appears 
around the AHC-hWC transition, which has been demonstrated to be a continuous band-inversion transition~\cite{x53d-12s6}. This completely agrees with our argument above, illustrated by Fig.~\ref{mu_Delta_pd}.

A natural question is why SDC only preempts the AHC-hWC transition, not the WC-AHC transition. The short answer is that when enforcing the commensurate condition ($\nu=1$), the AHC-hWC transition is indeed a continuous band-inversion transition, while the WC-AHC transition is first-order, in which case our argument above does not apply.
While the distinct nature of these two transitions is shown in Fig.~\ref{LJ_schematic}(d) through a SCHF calculation, we have understood that such an observation is not accidental.
It is, in fact, a consequence of the $C_6$ symmetry-protected topological (SPT) classification for these parent \emph{undoped} (\emph{commensurate}) crystals, which is well-defined beyond mean-field~\cite{fang2012bulk, po2017symmetry, song2017topological, thorngren2018gauging, song2020real}. 
Below, we show that the symmetry indicators for these three crystals are uniquely determined by the mechanism stabilizing them and can be analytically predicted from the quantum geometry of the parent continuum band. This chain of connections also demonstrates the central role of quantum geometry in stabilizing the SDC.

\paragraph*{Symmetry index of parent undoped crystals understood from parent band quantum geometry---} The jellium model has $C_6$ symmetry. In the following, we make an energetics assumption that the $C_6$ symmetry is never spontaneously broken. This means the crystalline state can be classified by $C_6$ symmetry indicators $(l_\Gamma, l_K, l_M)$. In Fig.~\ref{LJ_schematic}(b), we show a phase diagram for parent undoped crystals and their symmetry indices, which we now 
explain with analytical arguments. Our starting point is the WC at small $\lambda$, where the Berry curvature is dilute in $k$-space. This limit recovers the conventional jellium model. Therefore, we conclude that the symmetry index for WC is trivial $(l_\Gamma, l_K, l_M) = (0,0,0)$. As $\lambda$ rises, the Berry curvature becomes concentrated, driving the WC to transitions into an AHC with $C=1$. Important insight can be obtained by considering relatively weak coupling. The electrons mostly reside in the first Brillouin zone~(BZ$^1$), which is the regime for the ``Berry phase rounding" argument ~\cite{dong2024stability, soejima2024anomalous} to work. Namely, the crystal's Chern number is well estimated by
\begin{equation}
    C = \left\lfloor\frac{\phi({\rm BZ}^1)}{2\pi}\right\rceil
\end{equation}
As such, the boundary between WC and AHC is estimated by the condition $\phi({\rm BZ}^1) = \pi$. Crucially, this rounding argument reveals a physical picture of the WC-AHC transition: the boundary of ${\rm BZ}^1$ is analogous to a superconducting ring, whose vorticity jumps by one unit when the Berry curvature through the ring, i.e., the $k$-space analog of magnetic flux, approaches $2\pi$ -- the famous Little-Parks effect. An immediate implication from this picture is that only the BZ$^1$ boundary changes at this transition, while the $\Gamma$ remains unaffected, which means $l_\Gamma=0$ for AHC. Another constraint comes from the Chern number, which is related to $C_6$ symmetry indicators through
\begin{equation}
     C \equiv l_\Gamma + 2l_K + 3l_M \; ({\rm mod}\; 6)
     \label{C6_Chern}
\end{equation}
This eliminates all but one possible combination for the AHC with $C=1$: $(l_\Gamma, l_K, l_M) = (0, -1, 1)$.

The symmetry index for the hWC can also be inferred from the physical picture driving the AHC-hWC transition, which we understand as a consequence of the conspiracy between quantum geometry and interaction. The hWC appears at larger $\lambda$ and higher $r_s$. In this regime, one crucial condition breaks down: the electrons stop being confined to the first BZ by dispersion. Instead, the Fock energy, which 
disfavors electron occupation in regions with concentrated Berry curvature,  significantly renormalizes the dispersion into a Mexican hat-like shape~\cite{x53d-12s6}. As the electron gets pushed into higher zones, the Bloch function at $\Gamma$ can be approximated by a linear combination of plane waves corresponding to the six smallest non-zero reciprocal lattice vectors $G_i$. The transition regime is estimated by $\Delta E = E(G)-E(\Gamma)=0$, where $E(k)$ is the HF-renormalized dispersion.
Crossing this line from below, we expect the indices $l_K$ and $l_M$ to inherit that of the AHC, while $l_\Gamma$ is changed. In this situation, Eq.~\eqref{C6_Chern} allows us to determine the fate of $l_\Gamma$ by finding the Chern number.

To understand the Chern number, we propose an empirical ``generalized Berry curvature rounding" argument~\footnote{Unlike the argument for flux rounding in the first BZ, this argument is less controlled and more empirical. We note that a similar argument was numerically tested to work well for a series of toy models~\cite{Desrochers2026}.}. The Chern number is estimated from $\phi({\rm BZ}^2)$, the Berry curvature in the second BZ, where most electrons now live. Note that the total Berry curvature in the parent band of $\lambda$-jellium is $2\pi$, which means $\phi({\rm BZ}^2)<\pi$ as long as $\phi({\rm BZ}^1)>\pi$. The latter condition is always met when we increase $\lambda$ from the AHC phase. As such, one concludes $C=0$ and finds the symmetry index for hWC: $(l_\Gamma, l_K, l_M) = (-1, -1, 1)$. 

\paragraph*{Symmetry-forbidden continuous band-inversion transitions ---} The symmetry index implies that a direct WC-AHC transition cannot be continuous without fine-tuning. To see this, we track the change of index
\begin{equation}
    {\rm WC}: (0,0,0)\rightarrow {\rm AHC}:(0,-1,1)
\end{equation}
The indices at $K$ and $M$ both need to change. But there is no symmetry reason for the Dirac mass to vanish simultaneously at $K$ and $M$, meaning this transition, if continuous, must be two-step. Meanwhile, we also argue this transition is unlikely to be two-step, which would require an intermediate phase with $C=3$ or $C=-2$, not supported by the rounding argument. In other words, a direct WC-AHC transition is generically first-order.

In contrast, the transition between AHC and hWC
\begin{equation}
    {\rm AHC}:(0,-1,1) \rightarrow {\rm hWC}:(-1,-1,1)
\end{equation}
only toggles the $l_\Gamma$ index, and can be achieved by a Dirac band-inversion at $\Gamma$, which is indeed observed in earlier SCHF studies~\cite{soejima2024anomalous}.  

\begin{figure*}[t]
\centering
\includegraphics[width=0.99\textwidth]{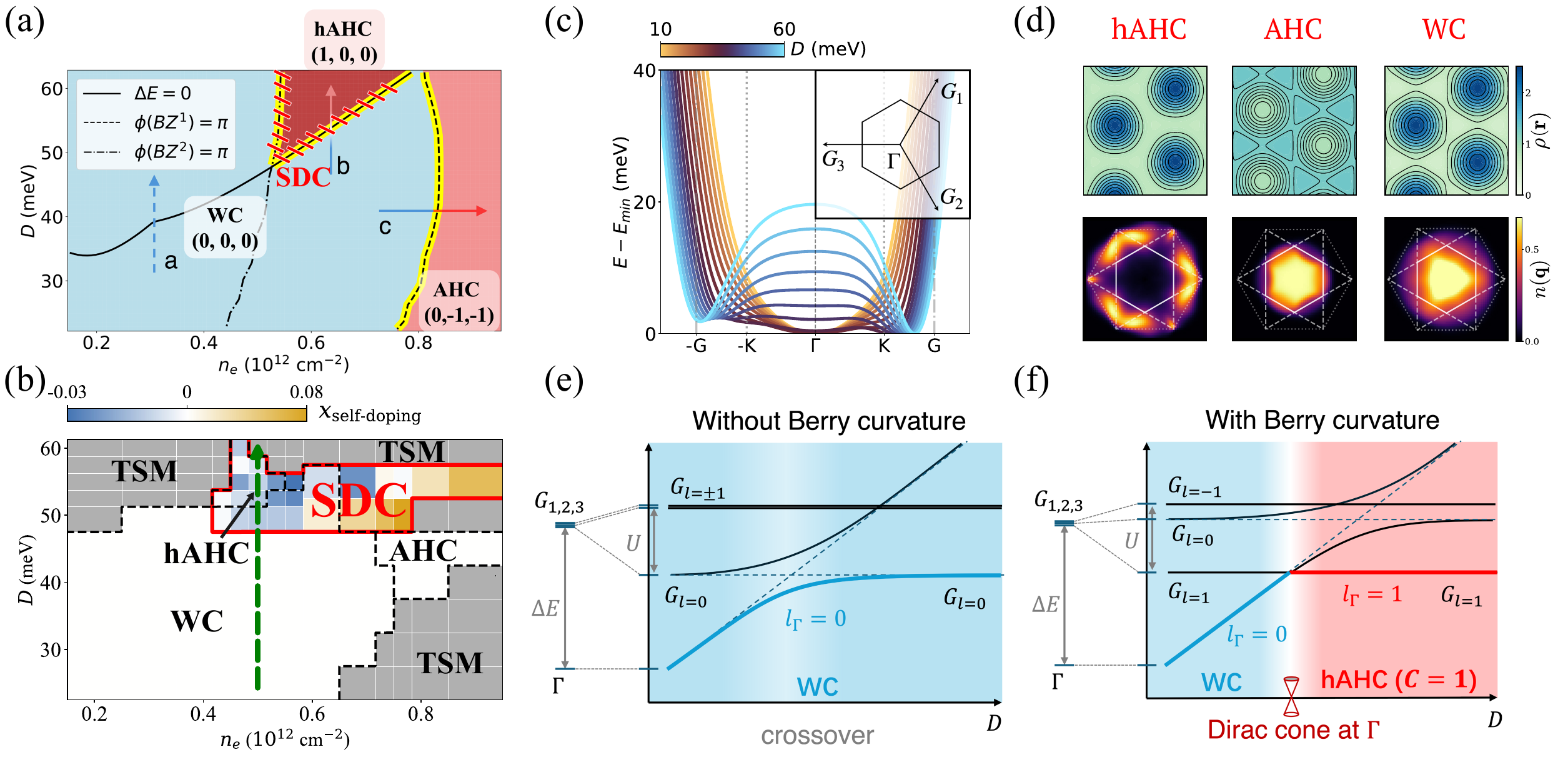}
\caption{Phase diagram for rhombohedral pentalayer graphene (R5G). (a)~Phase diagram for competing commensurate ($\nu=1$) crystalline orders predicted from band-structure criteria. The phase boundaries highlighted in yellow are estimated using the realistic model for R5G. Labels show $C_3$ symmetry indicators ($l_\Gamma, l_K, l_{K'}$) for three crystals. Red hatching marks the SDC we expect to preempt the WC-hAHC transition. Typical paths of evolution are marked by arrows in different colors, indicating their distinct nature: crossover, continuous transition, and first-order transition for paths a, b, and c, respectively.  
(b)~HF phase diagram obtained by enforcing commensurate order ($\nu=1$). Phase boundaries between commensurate crystals are marked with dashed lines. Color shows the estimated self-doping fraction. Grey denotes translation-symmetric metal (TSM), whose identification is detailed in the End Matter.  The green dotted line marks the line cut taken in Fig.~\ref{Fig:line-cut-r5g}(a).
(c) Evolution of R5G band structure as displacement field increases.
(d) Characterization of three crystals through real space density profile $\rho(\boldsymbol{r})$ and $k$-space occupation in parent band $n(\boldsymbol{q})$. 
(e)(f) Schematic diagrams showing the evolution of the Bloch function at $\Gamma$ as electrons migrate into the higher BZs, which correspond respectively to arrows $a$ and $b$ in panel (a) and are explained in detail in the main text. Background color density schematically shows the band gap at $\Gamma$.} 
\label{Fig:r5g_phase_diagram}
\end{figure*}

\paragraph*{Rhombohedral multilayer graphene: competition of undoped (commensurate) crystalline phases---}
As in our analysis above for the $\lambda$-jellium model, in light of the close connection between band-inversion transitions and self-doping, we first clarify the competition between parent states, i.e. the \textit{undoped} crystals, before addressing self-doping. The phase diagram of RMG has been extensively studied experimentally~\cite{lu2024fractional, han2024correlated,liu2024spontaneous,zhou2024layer,sha2024observation,han2024large,lu2025extended,choi2025superconductivity,xie2025tunable,lu2014quantum,ding2026spin,han2025signatures} and theoretically~\cite{dong2024theory, PhysRevLett.133.206503, dong2024stability, soejima2024anomalous, herzog2024moire, kwan2025moire}. To the best of our knowledge, in  existing studies on RMG the undoped crystals are classified only by Chern number or Hall conductance, and only the distinction between WC ($C=0$) and AHC ($C=1$) has been resolved. 
However, these crystalline phases also preserve $C_3$ rotation, which can protect distinctions between crystals with the same Chern number. This has recently been studied for crystals in a class of toy models~\cite{x53d-12s6, Desrochers2026}.

As is evident from the previous example, it is essential to resolve the symmetry-protected topological order of crystalline states. One byproduct of our study is showing that in the experimentally accessible regime and restricting to undoped (commensurate) crystals, there are in fact two distinct $C_3$-symmetric crystalline phases with $C=1$. Based on the symmetry indicator $l_\Gamma$, we label them AHC ($l_\Gamma=0$) and hAHC ($l_\Gamma = 1$). Their distinction is also evident from their real space density profile and $k$-space distribution, shown in Fig.~\ref{Fig:r5g_phase_diagram}(d). The difference between these two is similar to that between WC of $s$ orbitals and $p+ip$ orbitals (i.e. all symmetry indicators shifting by 1); the latter is also referred to as halo Wigner crystal (hWC)~\cite{x53d-12s6}.

Our theoretical prediction for the R5G phase diagram for \textit{undoped} (\textit{commensurate}) crystalline orders, based only on the discussed band-structure criteria (e.g. no SCHF) is shown in Fig.~\ref{Fig:r5g_phase_diagram}(a), which we will explain in full detail later. Its validity is supported by SCHF calculations in Fig.~\ref{Fig:r5g_phase_diagram}(b) over the same parameter range.
We note that the region where SDC is expected from our analytical arguments (indicated by red hatch marks in Fig.~\ref{Fig:r5g_phase_diagram}(a)) is narrow compared to the SCHF result.
\footnote{Meanwhile, the parameter range where the SDC is observed in Ref.~\cite{han2026evidence} is also narrower than the SCHF result, and, in fact, more consistent with our theoretical expectation.}
In addition, although we find an hAHC state in SCHF when enforcing commensuration $\nu=1$, this hAHC region is completely replaced by the SDC once we allow for self-doping. Hence, we refer to the parent commensurate crystal as a ``disqualified'' hAHC, following the title of Ref.~\cite{dong2026crystals}. 
This is also consistent with the experimental observation, which suggests no undoped crystal with $C=1$ in this parameter range.
In the following, we will deduce the undoped (commensurate) phase diagram from non-interacting band structures. This includes predicting all the $C_3$ indicators, hence the novel hAHC is also anticipated. For this purpose, it is crucial to understand the mechanism driving the phase transition between undoped crystals.

We first explain three crystal phases and their locations in the phase diagram. The key criteria involve the Berry flux through the BZs and the energy competition between $\Gamma$ and $G$ point. 
We start from the regime of low displacement field, in which case electrons are almost confined to BZ$^1$ by the dispersion. This is the regime where the ``Berry phase rounding" argument~\cite{dong2024stability, soejima2024anomalous} applies. At low density, a Wigner crystal is stabilized. As the density increases, more Berry flux enters BZ$^1$, and we expect an AHC with $C=1$. This part of discussion is completely parallel to the WC-AHC transition in $\lambda$-jellium discussed earlier.
The symmetry indicator of the AHC at low $D$ is easily predictable. In AHC, $l_\Gamma$ must be 0 since kinetic energy at small $D$ enforces occupation on $k=0$. Another constraint comes from the energetical assumption that $C_2$ symmetry is only weakly broken by the trigonal warping, which is consistent with what we observe from the real-space density profile. This leads us to conjecture $l_K=l_{K'}$, which leaves only one way to get $C=1$, that is, $(l_\Gamma, l_K, l_{K'}) = (0,-1,-1)$. This is confirmed in our SCHF calculation.

The scenario changes as we enhance displacement field, the continuum band structure deforms and develops a ``Mexican hat''  (see Fig.~\ref{Fig:r5g_phase_diagram}(c)). As electrons migrate to higher BZs, the rounding argument above based on $\phi({\rm BZ}^1)$ stops being relevant. This turning point is approximately captured by the criterion $\Delta E = E(G)-E(\Gamma)=0$ (see solid curve in Fig.~\ref{Fig:r5g_phase_diagram}(a)); here we crudely use the non-interacting dispersion, though a more quantitative estimate would include HF-renormalization. 
Above this line, we instead estimate the new phase boundary between $C=0$ and $C=1$ crystals using the generalized rounding condition based on the Berry flux $\phi({\rm BZ}^2)$ in higher BZs, as we did for the $\lambda$-jellium.
This crucially suggests that at higher $D$, the transition from $C=0$ crystal to $C=1$ crystal happens at a lower density, which creates space in the phase diagram for a new crystal to develop.

To understand this new crystal and determine its symmetry index, it is helpful to consider the evolution from WC as displacement field rises. Two typical scenarios correspoinding to arrows $a$ and $b$ in Fig.~\ref{Fig:r5g_phase_diagram}(a) are analyzed in Fig.~\ref{Fig:r5g_phase_diagram}(e) and (f). In both cases, the evolution is driven by a drastic change of the Bloch function at $\Gamma$ across the curve $\Delta E = E(G)-E(\Gamma)=0$, and we assume that the wavefunction at $K$ and $K'$ points are virtually untouched. At low density, electrons do not experience enough Berry curvature ($\phi({\rm BZ}^1),\phi({\rm BZ}^2)<\pi$), so that path $a$ ends up in the same WC phase, thus the index $l_\Gamma$ must remain 0. In contrast, at higher density, the electrons are occupying regions with strong Berry curvature. Then, the Berry flux rounding argument demands that path $b$ lands in a $C=1$ crystal. This suggests that the symmetry indicator $l_\Gamma$ must switch from 0 in WC to 1. This is how, in Fig.~\ref{Fig:r5g_phase_diagram}(f), we have conjectured that interaction must have split the degeneracy between three plane waves $\ket{G_{1,2,3}}$ into angular momentum eigenstates $\ket{G_{l=0,1,-1}}$ such that $l=1$ is lowest and crosses with $\ket{\Gamma}$ first. As such, we conclude that the $C=1$ crystal at higher $D$ is a halo anomalous Hall crystal (hAHC) with symmetry indicator $(l_\Gamma, l_K, l_{K'})=(1,0,0)$, which is confirmed by our SCHF \footnote{There are three $C_3$ Wyckoff positions, which means the $C_3$ indices are ambiguous for a crystal. However, we make an energetics assumption that throughout the phase diagram the $C_2$ is only weakly broken. This approximate $C_2$ selects one unique Wyckoff position, which we fix as the origin to uniquely determine the $C_3$ indices of crystals}.
Finally, we remark that the contrast between Fig.~\ref{Fig:r5g_phase_diagram}(e) and (f) also demonstrates the essential role of Berry curvature in stabilizing hAHC, which we will now explain is crucial to stabilizing SDC.

\begin{figure}[h]
\centering
\includegraphics[width=.99\columnwidth]{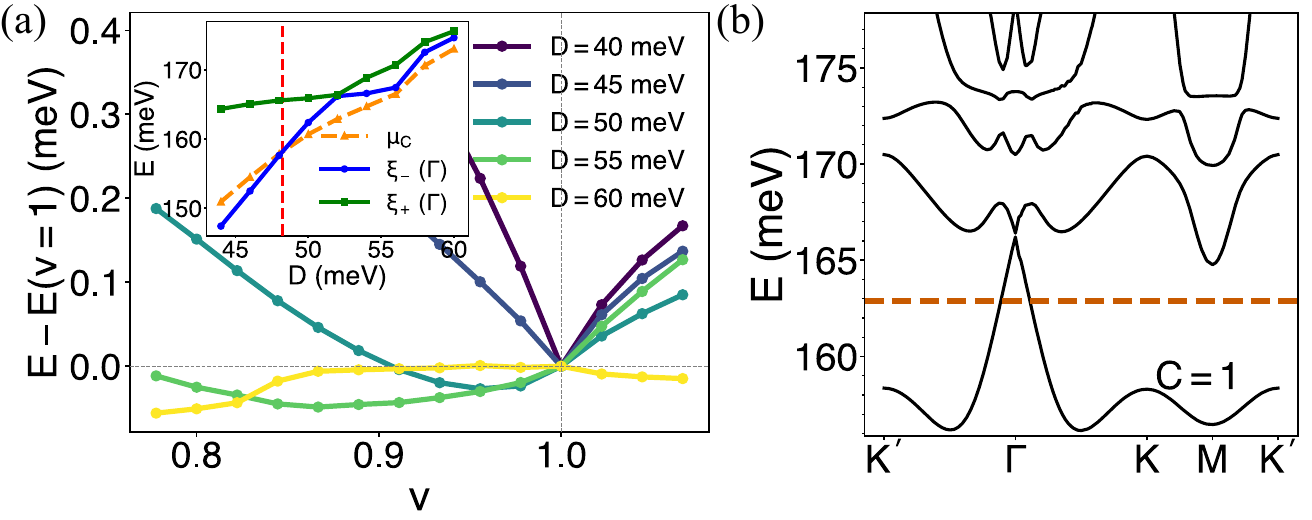}
\caption{Characterization of SDC phase of R5G at $n=0.5\times 10^{12}$cm$^{-2}$. (a) Ground-state energy difference relative to the commensurate ($\nu=1$) state as a function of $\nu$. Different colors denote different displacement field $D$ strength. Inset: evolution along the green linecut shown in Fig.~\ref{Fig:r5g_phase_diagram}(b), comparison between valence and conduction bands energy at $\Gamma$, and chemical potential of $\nu=1$ state. Self-doping of holes appears at $D\approx 48 $meV while band inversion happens at $D\approx 52 $meV. (b) Hartree-Fock band structure for $\nu=1$ state for $D=52$ meV. The orange dashed line denotes the corresponding chemical potential $\mu_C$.}
\label{Fig:line-cut-r5g}
\end{figure}

\paragraph*{WC-hAHC transition preempted by SDC---} Having clarified the symmetry index of three crystals, we are ready to discuss the transition between them. The presence of this hAHC is essential to a gap-closing transition, hence crucial to stabilizing SDC. This is because the symmetry indicators for WC and hAHC
\begin{equation}
    {\rm WC}: (0,0,0) \rightarrow {\rm hAHC}: (1,0,0)
\end{equation}
only differ by 1 on the $l_\Gamma$, which allows a Dirac band touching to bridge the two phases. In Fig.~\ref{Fig:line-cut-r5g}, we verify this Dirac transition in SCHF.

\begin{figure}[h]
\centering
\includegraphics[width=\columnwidth]{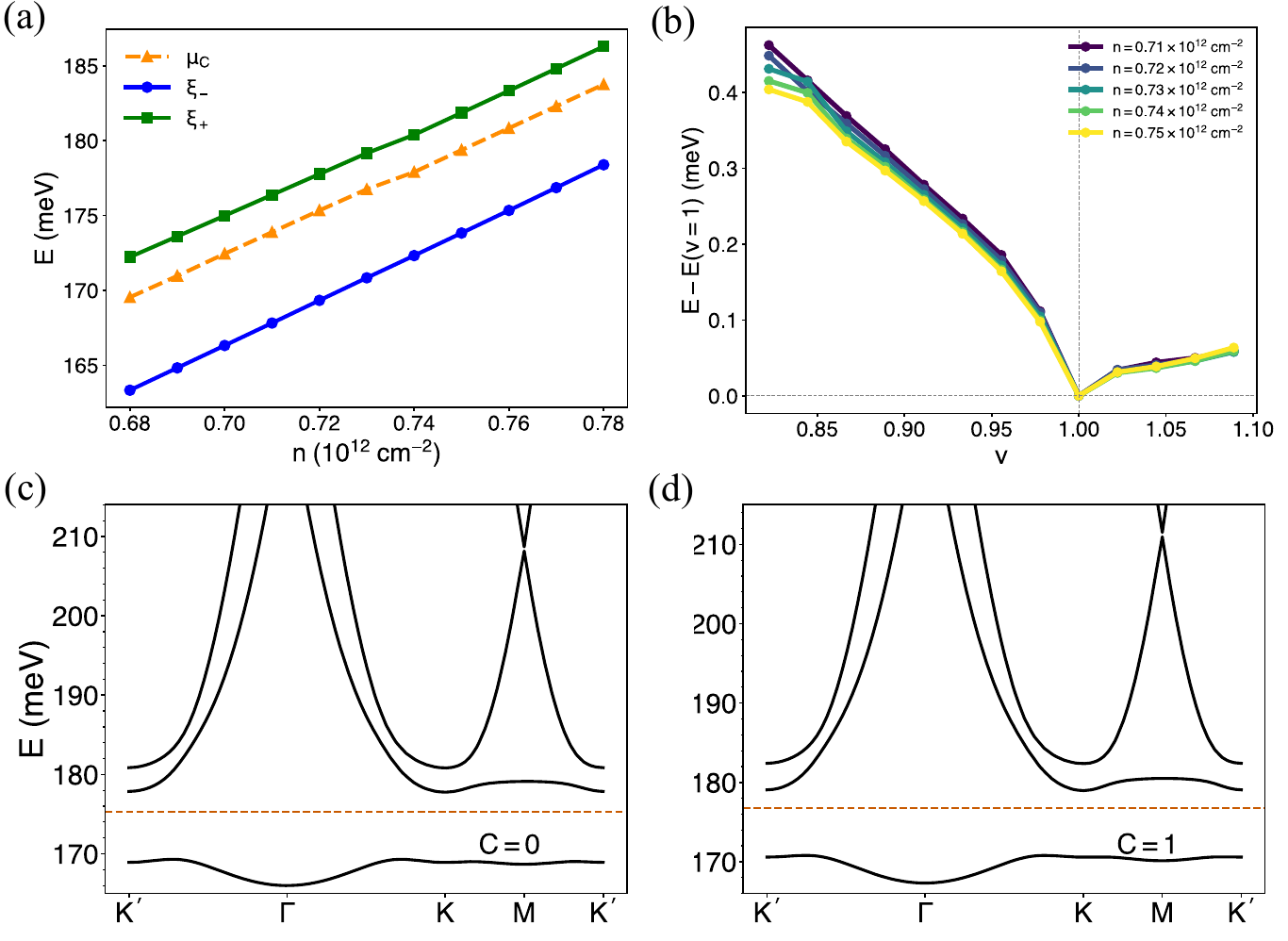}
\caption{Absence of SDC along the first-order WC-AHC transition in R5G. (a) The evolution of valence band top $\xi_-$ (blue circles), conduction band bottom $\xi_+$ (green squares), and chemical potential $\mu_C$ (orange triangles) of $\nu=1$ state as a function of $n$ under $D=40$ meV. (b) Ground-state energy difference relative to the commensurate ($\nu=1$) state as a function of $\nu$. HF band structures for $\nu=1$ state for (c) $n=0.73\times 10^{12}$cm$^{-2}$ and (d) $n=0.74\times 10^{12}$cm$^{-2}$. Orange dashed line denotes the chemical potential $\mu_C$.}
\label{r5g_linecut2}
\end{figure}

In contrast, a continuous evolution from WC to AHC
\begin{equation}
    {\rm WC}: (0,0,0) \rightarrow {\rm AHC}: (0,-1,-1)
\end{equation}
would require two Dirac band touchings at $K$ and $K'$, which already requires a two-step transition, as there is no $C_2$ symmetry enforcing the two band-inversions to be simultaneous. 
The problem list of the continuous assumption does not end there. If we make the general assumption~\footnote{As we are aware that the $C_3$ would also allow $k^2$ Dirac cones with opposite winding, but it would require fine-tuning to suppress the leading $O(k)$ terms. So in the most general scenario, we expect linear Dirac cones.} such that the critical points for band inversions at $K$ and $K'$ are both described by a linear Dirac cone, then, given the change of symmetry indicators, each band inversion would only change the crystal Chern number by $\Delta C = -1$. As a result, this two-step transition leads to a $C=-2$ crystal instead of the $C=1$ AHC we wanted. To fix this within the continuous assumption, another transition is needed where three Dirac cones on non-high-symmetry momentum related by $C_3$ go through band inversion together. This would change the Chern number by 3 while keeping the symmetry indicators. But then the WC-AHC transition would become a three-step transition. This picture naturally reduces to the $\lambda$-jellium example we discussed earlier when a $C_2$ symmetry is imposed, as $C_2$ pins the 3 Dirac cones to $M$ points and synchronizes the two transitions for band inversion at $K$ and $K'$, reducing it into a two-step transition. Therefore, we conclude that a direct WC-AHC transition must be discontinuous. This is verified by our SCHF result in End Matter Fig.~\ref{r5g_linecut2}. 
Finally, combining with our argument in Fig.~\ref{mu_Delta_pd}, we expect the SDC to appear only in the vicinity of the phase boundary between WC and hAHC, which is consistent with our SCHF result in Fig.~\ref{Fig:r5g_phase_diagram}(b) as well as the experiments.

\paragraph*{Conclusion---} To address the recent observation of self-doped crystal (SDC) in rhombohedral multilayer graphene, our work establishes the following picture: a SDC is generically stabilized from a preempted band-inversion transition between two undoped crystals. The possibility of a band-inversion transition relies on matching symmetry indicators of two crystals, which are predictable from the parent band quantum geometry. With the self-consistent Hartree-Fock method, we find self-doped crystals in two examples: the $\lambda$-jellium model and rhombohedral pentalayer graphene, and demonstrate that our theoretical understanding reliably predicts the presence of SDC.

\paragraph*{Acknowledgment---} We thank Tianle Wang, Taige Wang, Pavel Nosov, Felix Desrochers, Yahui Zhang, and particularly Tonghang Han and Patrick Ledwith for helpful discussions. We thank Junkai Dong, Tomohiro Soejima, Daniel Parker, and Ashvin Vishwanath for communications regarding their recent independent work~\cite{dong2026crystals} on a related topic, including their choice of parameters. Although our works both concern the topic of self-doped crystals in RMG, the focuses and perspectives are distinct, and the overlapping parts of our results agree. 
This work was primarily funded by the U.S. Department of Energy, Office of Science, Office of Basic Energy Sciences, Materials Sciences and Engineering Division under Contract No. DE-AC02-05-CH11231 (Theory of Materials program KC2301).
Z.~H. is supported by a Simons Investigator award, the Simons Collaboration on Ultra-Quantum Matter, which is a grant from the Simons Foundation (651440, Ashvin Vishwanath).

\bibliographystyle{apsrev4-2}
\bibliography{iwc}

\onecolumngrid
\newpage
\twocolumngrid
\section*{End matter}

\begin{figure}[h]
\centering
\includegraphics[width=\columnwidth]{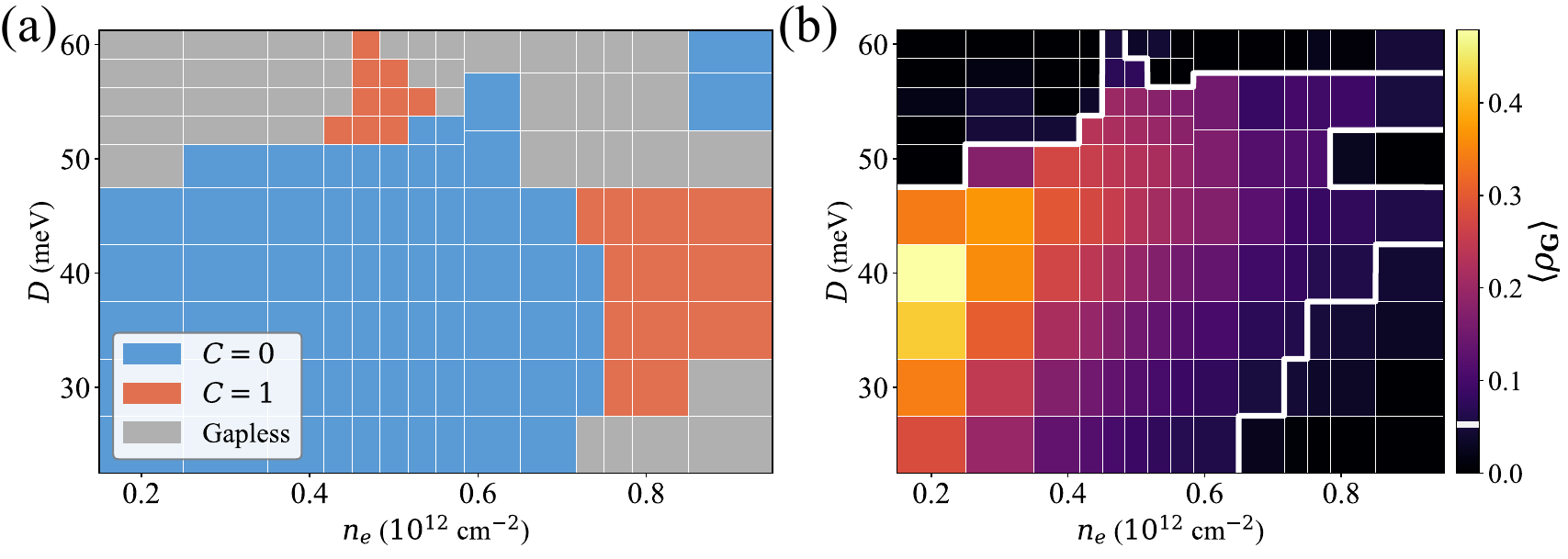}
\caption{(a) HF phase diagram for R5G, obtained by a HF ansatz with commensurate filling ($\nu=1$). The Chern number of the lowest HF band is marked, with blue for $C=0$ and red for $C=1$. Grey denotes regions where the direct HF band gap is under 1 meV. (b) Characterization of continuous translation symmetry breaking in R5G in the same parameter region as Fig.~\ref{Fig:r5g_phase_diagram}(a) and (b). Color shows the average of $\langle \rho_{\mathbf{G}} \rangle$ among three first-shell reciprocal lattice vectors $\{ G_1, G_2, G_3\}$. The white line marks the boundary of crystalline regions, defined by a threshold $\langle\rho_\mathbf{G}\rangle>0.05$. This SCHF calculation is done with a commensurate CDW ordering momentum $G$ corresponding to $\nu=1$.}
\label{Fig:r5g_CDWorder}
\end{figure}

\subsection*{Details on thermodynamic criterion for self-doping}
In the following, we properly set up the problem and derive the intuitive results in the main text. Consider a system of electrons with density $n_e$ that forms a crystalline order with lattice density $n_c$. We define the electron filling per unit cell $\nu=\frac{n_e}{n_c}$. For example, $\nu=1$ corresponds to undoped crystals. 

Next, we restrict to a continuous transition from WC to SDC. Namely, the ground-state lattice density $n_c$ is continuously drifting away from $n_e$.
A commensurate crystalline order gets destabilized when the energy density
\begin{equation}
    \epsilon=\epsilon(n_c, \nu) = \lim_{A\rightarrow \infty}{\frac{E(n_c, \nu, A)}{A}} \label{limit}
\end{equation}
stops being minimum at $\nu=1$, namely
\begin{align}
    \left(\frac{\partial \epsilon}{\partial \nu}\right)_{n_e}\bigg|_{\nu=1^-} < 0  &\Rightarrow \text{hole-doped SDC}\nonumber \\
    \left(\frac{\partial \epsilon}{\partial \nu}\right)_{n_e}\bigg|_{\nu=1^+} > 0  &\Rightarrow \text{electron-doped SDC}
\end{align}
These two limits approaching $\nu=1$ differ by the charge gap. We note that this is a sufficient but not necessary condition to get SDC, since $\epsilon(\nu)$ may have multiple local minima. To show what this condition implies, it is useful to express $\left(\frac{\partial \epsilon}{\partial \nu}\right)_{n_e}$ in terms of thermodynamic quantities for WC (at fixed $\nu=1$). Using
\begin{align}
    d\epsilon &= \left(\frac{\partial \epsilon}{\partial n_c}\right)_{\nu} dn_c + \left(\frac{\partial \epsilon}{\partial \nu}\right)_{n_c} d\nu\\
    n_c &= n_e/\nu
\end{align}
we find
\begin{align}
    \left(\frac{\partial \epsilon}{\partial \nu}\right)_{n_e}\bigg|_{\nu=1^\pm} = \left(\frac{\partial \epsilon}{\partial n_c}\right)_{\nu} \left(\frac{\partial n_c}{\partial \nu}\right)_{n_e} + \left(\frac{\partial \epsilon}{\partial \nu}\right)_{n_c}\nonumber \\
    = -\frac{n_e \mu_{C}}{\nu} + n_c \xi_{\pm} = n_c (\xi_\pm -\mu_{C})\label{criterion_em}
\end{align}
where we have defined the chemical potential for the WC
\begin{align}
    \mu_{C} = \left(\frac{\partial \epsilon}{\partial n_e}\right)_{\nu=1}= \left(\frac{\partial \epsilon}{\partial n_c}\right)_{\nu=1} \label{muWC}
\end{align}
and band edge of doped electron($\xi_+$)/hole($\xi_-$)
\begin{align}
    \xi_\pm = \xi(n_c, \nu=1^\pm) = \left(\frac{\partial \epsilon}{\partial n_e}\right)_{n_c}\bigg|_{\nu=1^\pm} = \frac{1}{n_c}\left(\frac{\partial \epsilon}{\partial \nu}\right)_{n_c}\bigg|_{\nu=1^\pm}
\end{align}
Remarkably, both quantities can be determined by studying an undoped crystal at fixed $\nu=1$. Therefore, Eq.~\eqref{criterion} enables us to predict the instability toward SDC completely from the properties of an undoped crystal.

\newpage
\onecolumngrid

\appendix

\section{Modeling of rhombohedral multilayer graphene}

This section describes the details of the microscopic model of rhombohedral multilayer graphene.
The Hamiltonian is given by~\cite{PhysRevLett.133.206503,PhysRevB.82.035409,PhysRevB.88.075408}
\begin{equation}
    \hat{H} = \hat{h}_{\text{kin}} + \sum_{\boldsymbol{q}} \frac{U_{|\boldsymbol{q}|}}{2A} :\!\hat{\rho}_{\boldsymbol{q}} \hat{\rho}_{-\boldsymbol{q}}\!:, \quad U_q = \frac{2\pi \tanh(qd)}{\epsilon_r \epsilon_0 q},
\end{equation}
where $A$ is the sample area and normal ordering is relative to the charge neutrality gap of RMG.
Screened Coulomb potential has gate distance $d=250\AA$ and dielectric constant $\epsilon_r=15$.
Notably, $\epsilon_r$ is an important parameter in the model which controls the strength of Coulomb interaction.
The kinetic term takes the form
\begin{equation}
    h_{\rm kin} = h_{\rm RG}^{(N_L)} + h_D ,
\end{equation}
where $h_{\rm RG}^{(N_L)}$ is a standard model for $N_L$-layer rhombohedral graphene with intralayer and interlayer hoppings $t_{0-4}$~\cite{zhou2021half}.
We used $N_L=5$ for all our calculations.
For readers' convenience, we reproduce its form here~\cite{soejima2024anomalous,PhysRevLett.132.116504}:
\begin{equation}
    h_{\text{RG}}^{(N_L)}(\mathbf{k}) = \begin{pmatrix}
h_\ell^{(0)} & h^{(1)} & h^{(2)} & & & \\
h^{(1)\dagger} & h_\ell^{(0)} & h^{(1)} & h^{(2)} & & \\
h^{(2)\dagger} & h^{(1)\dagger} & h_\ell^{(0)} & \ddots & \ddots & \\
& h^{(2)\dagger} & \ddots & \ddots & h^{(1)} & h^{(2)} \\
& & & h^{(1)\dagger} & h_\ell^{(0)} & h^{(1)} \\
& & & h^{(2)\dagger} & h^{(1)\dagger} & h_\ell^{(0)}
\end{pmatrix},
\end{equation}
which is a matrix with entries in the sublattice space
\begin{subequations}
\begin{align}
h_\ell^{(0)}(\mathbf{k}) &= \begin{pmatrix}
u_{A\ell} & -t_0 f_{\mathbf{k}} \\
-t_0 f_{\mathbf{k}}^* & u_{B\ell}
\end{pmatrix} \label{eq:h0} \\
h^{(1)}(\mathbf{k}) &= \begin{pmatrix}
t_4 f_{\mathbf{k}} & t_3 f_{\mathbf{k}}^* \\
t_1 & t_4 f_{\mathbf{k}}
\end{pmatrix} \label{eq:h1} \\
h^{(2)}(\mathbf{k}) &= \begin{pmatrix}
0 & \frac{t_2}{2} \\
0 & 0
\end{pmatrix} \label{eq:h2} \\
f_{\mathbf{k}} &= \sum_{i=0}^{2} e^{i \mathbf{k} \cdot \boldsymbol{\delta}_i} \label{eq:fk}.
\end{align} 
\end{subequations}
We use parameters $(t_0,t_1,t_2,t_3,t_4)=(3100,380,-21,290,141)$meV.
On-site potentials $h_D$ are given by the diagonal term
\begin{equation}
    h_D^{\sigma,\ell} = u_D \left( \ell + 1 - \frac{N_L - 1}{2} \right) + \Delta' \left( 1 - \delta_{\sigma A}\,\delta_{\ell,1} - \delta_{\sigma B}\,\delta_{\ell,N_L} \right) ,
\end{equation}
where $\Delta'=10.5$meV represents the dimer site crystal field splitting, and $u_D>0$ is the externel displacement field.
Specifically, we take the moiré potential strength $V_{\rm hBN} \to 0$.
This makes our setting simpler and excludes the effects of moiré potential.

Apart from $u_D$, another important parameter for our RMG model is electron density $n_e$.
Under commensurate filling $\nu=1$, we have $n_e=n_c$, so we control $n_e$ by setting the corresponding $n_c$.
We define unit cell vectors $\mathbf{L}_1=a(1/2,\sqrt{3}/2)$, $\mathbf{L}_2=a(1/2,-\sqrt{3}/2)$.
The unit cell area is then $\sqrt{3} a^2/2$.
Under SI units, the relationship between $n_c$ and $a$ is then
\begin{equation}
    a = \sqrt{\frac{200}{\sqrt{3} n_c}} {\rm nm} ,
\end{equation}
where we assume $n_c$ is in unit of $10^{12}$cm$^{-2}$.
We define reciprocal lattice vectors by $\mathbf{g}_i\cdot \mathbf{L} _j = 2\pi \delta_{ij}$.
We define high-symmetry points at
\begin{equation}
    \Gamma = \mathbf{K}_{\text{Gr}}, \quad K = \Gamma - \frac{1}{3}\mathbf{g}_1 + \frac{1}{3}\mathbf{g}_2, \quad K' = \Gamma - \frac{2}{3}\mathbf{g}_1 - \frac{1}{3}\mathbf{g}_2, \quad M = \Gamma - \frac{1}{2}\mathbf{g}_1.
\end{equation}

In order to truncate the Hilbert space, we keep $N_b=10$ bands above charge neutrality for a self-consistent Hartree-Fock calculation.
We use Monkhorst-Pack grids with $30\times 30$ unit cells.
For each point reported in the main text, we use 10 random seeds for initialization and regard the one with the lowest energy as the converged ground state.

\section{Modeling of $\lambda$-Jellium model}

We start from the Hamiltonian introduced in Ref.~\cite{x53d-12s6}.
Different from the original form, we assume length unit is $a_0$ while energy unit is still Ry.
The Hamiltonian we used for our SCHF calculation can be written as
\begin{equation}
    \hat{H} = \hat{h}_{\text{kin}} + \sum_{\boldsymbol{q}} \frac{V_q}{2A} :\!\hat{\rho}_{\boldsymbol{q}} \hat{\rho}_{-\boldsymbol{q}}\!: .
\end{equation}
The single particle part can be written as
\begin{equation}
    \hat{h}_{\text{kin}} =  \Delta \begin{pmatrix} (\lambda r_s)^2 q^2 & -(\lambda r_s) (q_x - i q_y) \\ -(\lambda r_s)(q_x + i q_y) & 1 \end{pmatrix} + \hat{I}_2 q^2  ,
\end{equation}
where $\lambda$ controls the concentration of Berry curvature in the lower band and $r_s$ is the familiar potential to kinetic ratio controlling the strength of interaction.
$\hat{I}_2$ denotes a unit matrix.
We set $\Delta$ to some large number so that the upper band is projected out.
The original interaction $V_q=4\pi/q$ contains a singularity.
In order to suppress the singular behavior at $q=0$, we use the screened Coulomb potential
\begin{equation}
    V_q = \frac{4\pi \tanh(qd)}{q},
\end{equation}
where we set $d=100 a_0$.
The large gate distance $d$ ensures the screening is weak, so it is only used to suppress the singularity of interaction.

In order to truncate the Hilbert space, we keep $N_b=7$ bands above charge neutrality for a self-consistent Hartree-Fock calculation.
We use Monkhorst-Pack grids with $30\times 30$ unit cells.
For each point reported in the main text, we use 10 random seeds for initialization.

\section{Self-consistent Hartree-Fock (SCHF) method}
\label{app:SCHF}

Here we describe the details of our self-consistent Hartree-Fock (SCHF) method.
We mainly elaborate on the numerics of R5G model, because it has more complexities (e.g. flavor symmetry, choices of graphene-scale parameter) than the $\lambda$-jellium model.
We start with single-particle density matrices $P(\mathbf{k})_{\alpha \beta}=\langle c^\dagger_{\mathbf{k} \alpha} c^\dagger_{\mathbf{k} \beta}\rangle$, where $\alpha$ and $\beta$ denote band indices.
We assume spin and valley-polarized states in all our numerics in this work, which is supported by experiments~\cite{zhou2021half,zhou2022isospin,zhang2023enhanced,han2023orbital} and SCHF numerics~\cite{chatterjee2022inter,PhysRevLett.132.116504} in relevant $D$ and $n_e$ regimes. 
Hartree and Fock Hamiltonians are defined as~\cite{PhysRevLett.133.206503}
\begin{subequations}
\begin{align}
h_H[P](\boldsymbol{k}) &= \frac{1}{A} \sum_{\boldsymbol{g}} V_{\boldsymbol{g}} \Lambda_{\boldsymbol{g}}(\boldsymbol{k}) \left( \sum_{\boldsymbol{k}'} \text{Tr}[P(\boldsymbol{k}') \Lambda_{\boldsymbol{g}}(\boldsymbol{k}')^\dagger] \right) , \\
h_F[P](\boldsymbol{k}) &= -\frac{1}{A} \sum_{\boldsymbol{q}} V_{\boldsymbol{q}} \Lambda_{\boldsymbol{q}}(\boldsymbol{k}) P([\boldsymbol{k} + \boldsymbol{q}]) \Lambda_{\boldsymbol{q}}(\boldsymbol{k})^\dagger .
\end{align}
\end{subequations}
where $[\Lambda_{\boldsymbol{q}}(\boldsymbol{k})]_{\alpha\beta} = \langle \psi_{\boldsymbol{k}\alpha} | e^{-i\boldsymbol{q} \cdot \boldsymbol{r}} | \psi_{\boldsymbol{k}+\boldsymbol{q}\beta} \rangle$ are form factors.
The sum over $\boldsymbol{g}$ runs over the reciprocal vectors and $\boldsymbol{q}$ runs over all momentum transfers.
The energy of the Slater-determinant state can be calculated via Wick's theorem~\cite{PhysRevLett.133.206503}
\begin{equation}
    E[P] = \frac{1}{2} \text{Tr}[P(2h_{\text{kin}} + h_H[P] + h_F[P])] ,
\end{equation}
where the trace is over momentum and all band labels.
We use the optimal damping algorithm (ODA) to converge to states satisfying the self-consistency condition
\begin{equation}
    [P, h_{\text{kin}} + h_H[P] + h_F[P]] = 0 ,
\end{equation}
with tolerances denoted as $E_{\rm tol}$.
We use $E_{\rm tol} = 10^{-9}$ meV for R5G model.
To avoid non-global minima, we initialize SCHF with many independent random states as stated earlier.

Next we briefly state how we differentiate different ground states in HF.
in Fig.~\ref{Fig:r5g_CDWorder}(a) we regard states with direct HF gap smaller than 1 meV as ``gapless'' states.
For those states with a direct HF gap, Chern number is well defined for the lowest band.
As stated in the main text, it is possible (and important) that gapless states might break translational symmetry.
Therefore, gapless states (also called ``metal'') can be classified into two phases: those preserve tranlational symmetry (what we call TSM in the main text), and those break translation symmetry (and hence self-dope and we call SDC).
In order to measure translation symmetry breaking, we calculate so-called CDW order parameter:
\begin{equation}
\label{eq:CDWorder}
    \langle \rho_{\mathbf{Q}} \rangle 
= \frac{1}{N_k} \sum_{\mathbf{k}} 
  \mathrm{Tr}\!\Big[\Lambda_{\mathbf{Q}}(\mathbf{k})\;P(\mathbf{k})\Big] ,
\end{equation}
where trace is over all band indices and $N_k$ is the number of total $k$-points.
We calculate Eq.~\eqref{eq:CDWorder} for $\mathbf{Q}= \{G_1, G_2, G_3 \}$ in Fig.~\ref{Fig:r5g_phase_diagram}(c), the first-shell reciprocal lattice vectors.
The average value of the three $\langle \rho_{\mathbf{Q}}\rangle$ values are plotted in Fig~\ref{Fig:r5g_CDWorder}(b), and we use $|\langle \rho_{\mathbf{Q}} \rangle|<0.05$ as the criterion for TSM in Fig.~\ref{Fig:r5g_phase_diagram}(b).


\end{document}